\documentclass[%
 aip,
 amsmath,amssymb,
 reprint,%
]{revtex4-1}

\usepackage{graphicx}
\usepackage{dcolumn}
\usepackage{bm}

\usepackage[utf8]{inputenc}
\usepackage[T1]{fontenc}
\usepackage{mathptmx}
\usepackage{color}
\usepackage[normalem]{ulem}

\usepackage{microtype}

\begin{document}

\preprint{AIP/123-QED}

\title{Perspective: Strongly correlated and topological states in [111] grown transition metal oxide thin films and heterostructures}

\author{Jak Chakhalian}
 \email{jak.chakhalian@rutgers.edu.}
 \affiliation{Department of Physics, Rutgers University, Piscataway, New Jersey 08854-8019, USA}
 
\author{Xiaoran Liu}
 \email{xiaoran.liu@rutgers.edu.}
 \affiliation{Department of Physics, Rutgers University, Piscataway, New Jersey 08854-8019, USA}

\author{Gregory A. Fiete}
\email{g.fiete@northeastern.edu.}
\affiliation{Department of Physics, Northeastern University, Boston, Massachusetts 02115, USA}
\affiliation{Department of Physics, Massachusetts Institute of Technology, Cambridge, Massachusetts 02139, USA}

\date{\today}

\begin{abstract}
We highlight recent advances in the theory, materials fabrication, and experimental characterization of strongly correlated and topological states in [111] oriented transition metal oxide thin films and heterostructures, which are notoriously difficult to realize compared to their [001] oriented counterparts.  We focus on two classes of complex oxides, with the chemical formula ABO$_3$ and A$_2$B$_2$O$_7$, where the B sites are occupied by an open-shell transition metal ion with a local moment, and the A sites are typically a rare earth. The [111] oriented quasi-two-dimensional lattices derived from these parent compound lattices can exhibit peculiar geometries and symmetries, namely, a buckled honeycomb lattice, as well as kagome and triangular lattices.  These lattice motifs  form the basis for emergent strongly correlated and topological states expressed in exotic magnetism, various forms of orbital ordering, topological insulators, topological semimetals, quantum anomalous Hall insulators, and quantum spin liquids. For transition metal ions with high atomic number, spin-orbit coupling plays  a significant role and may give rise to additional topological features in the  electronic band structure and in the spectrum of magnetic excitations.  We conclude the Perspective by articulating open challenges and opportunities in this actively  developing field. \end{abstract}

\maketitle

\section{Introduction}
\label{sec:intro}
Physicists undertake a continuous  quest for high quality quantum materials in the thin-film form that harbor interesting  many-body  phases and are tunable by substrate strain,  dimensional control, and compositional modulation.  Such systems typically allow for the realization of a complex many-body ground state with multiple entwined  correlated and topological states that may--in principle--be tuned as a function of external control parameters, such as external electric and magnetic fields, optical  excitations, gating, and pressure.   

Over the past decade, transition metal oxide interfaces have been a subject of intense investigation with the prime focus on the complex oxides in the ABO$_3$ perovskite form (here A is typically a rare earth element, B is an open-shell transition metal element, and O is oxygen). These thin films and heterostructures are predominantly grown along the [001] direction \cite{Mannhart:mrb08,Zubko:arcmp11,Ahn:rmp06,Chakhalian:prl11,Huang:am18,Boschker:arcmp17,Yadav:nat16,Chakhalian:rmp14}.  As a vivid illustration of the physical possibilities, some of the [001] oriented interfaces composed of two insulating materials show the emergence of a two-dimensional electron gas (2DEG) with electronic states derived primarily from the $d$-orbitals of the transition metal ions \cite{Pentcheva:prl07,Pentcheva:jpcm10,Rabe:arcmp10}, rather than the more extended and less weakly correlated $s,p$-orbitals common to semiconductor interfaces.  Compared to the extremely high mobility of carriers in GaAs/AlGaAs heterostructures used to study the fractional quantum Hall effect \cite{Stormer:rmp99}, the 2DEGs at the [001] oxide interfaces have significantly lower mobility but exhibit markedly stronger electron-electron correlations \cite{Millis:prb10,Ohtomo:nat04,Okamoto:nat04,Thiel:sci06,Chakhalian:sci07,Chakhalian:nm12,Hwang:nm12}  that enable a richer set of compelling phenomena including the quantum Hall effect \cite{Tsukazaki:sci07}, the fractional quantum Hall effect \cite{Tsukazaki:nm10}, unconventional superconductivity \cite{Reyren:sci07,Li:np11,Bert:np11}, and magnetism \cite{Brinkman:natmat07,Bert:np11}.  The large variety of accessible states in these systems can potentially usher in a new era of oxide electronics that exploit these correlated states \cite{Ramirez:sci07,Mannhart:sci10}.

In this Perspective, we focus on a new direction which is still in its infancy -- transition metal oxide films templated in high-index  (e.g. [111]) directions.  Compared to the [001] direction, growth along [111] is more difficult due to the lack of readily available lattice-matched substrates, and less favorable and generally little understood thermodynamics of such growth.  The first step in this direction was taken with the  ABO$_3$ perovskites, both on the theoretical \cite{Xiao:nc11,Ruegg:prb12,Ruegg_Top:prb13,Ruegg11_2,Wang:prb11} and experimental sides \cite{Blok:apl11,Herranz:sp2012,Middey:apl12}.  An important motivator was the theoretical prediction that correlated topological states, such as the quantum anomalous Hall state, could be realized in ultra-thin films of [111] oriented perovskites \cite{Fiete:jap15}, where the transition metal ions in a bilayer system form a buckled honeycomb lattice. This was an important step because the honeycomb lattice has played a pivotal role in the pioneering theoretical studies of topological states on lattices \cite{Kane:prl05,Kane_2:prl05,Haldane:prl88}.

While the perovskite systems are interesting in their own right, in this Perspective we  also  focus on the interesting class of compounds A$_2$B$_2$O$_7$ which brings an additional element of geometrical frustration, compared to the ABO$_3$ materials. In A$_2$B$_2$O$_7$ compounds, the B sites form a pyrochlore lattice--a network of corner sharing tetrahedra.   Viewed along the [111] direction, the lattice consists of alternating kagome and triangular lattice planes.  Both the kagome and triangular lattices lead to magnetic frustration, which can support exotic fractionalized states--a major motivation for experimentally achieving high quality materials growth for such systems. The three-dimensional pyrochlore lattice is fully frustrated as well, and can give rise to fractionalized topological insulators \cite{Maciejko:np15} and topologically  ordered magnetic states such as quantum spin liquids \cite{Balents:nat10}, which may portend what is possible with the thin-film versions  highlighted in this Perspective.

This Perspective is organized as follows.  In Sec.\ref{sec:perovskite} we describe several theoretical proposals for novel electronic states and experimental results in the [111] grown ABO$_3$ perovskite bilayer, which exhibits a buckled honeycomb lattice.  In Sec.\ref{sec:pyrochlore} we describe  theoretical proposals for novel electronic states and experimental results on [111] oriented A$_2$B$_2$O$_7$. We highlight the challenges of the A$_2$B$_2$O$_7$ pyrochlore growth compared to the heavily investigated films of the ABO$_3$ family. Finally, in Sec.\ref{sec:outlook}, we conclude with an outlook towards possible future routes  to investigation of these quantum materials.  

We note that we do not aim to present a comprehensive review of this fast moving field, but rather provide our perspective of the challenges and opportunities in this direction.  We apologize to those whose work is not prominently featured here or inadvertently omitted.

\section{[111] grown ABO$_3$ bilayer: A buckled honeycomb lattice}
\label{sec:perovskite}

\begin{figure}[ht]
\begin{center}
\includegraphics[width=.5\textwidth]{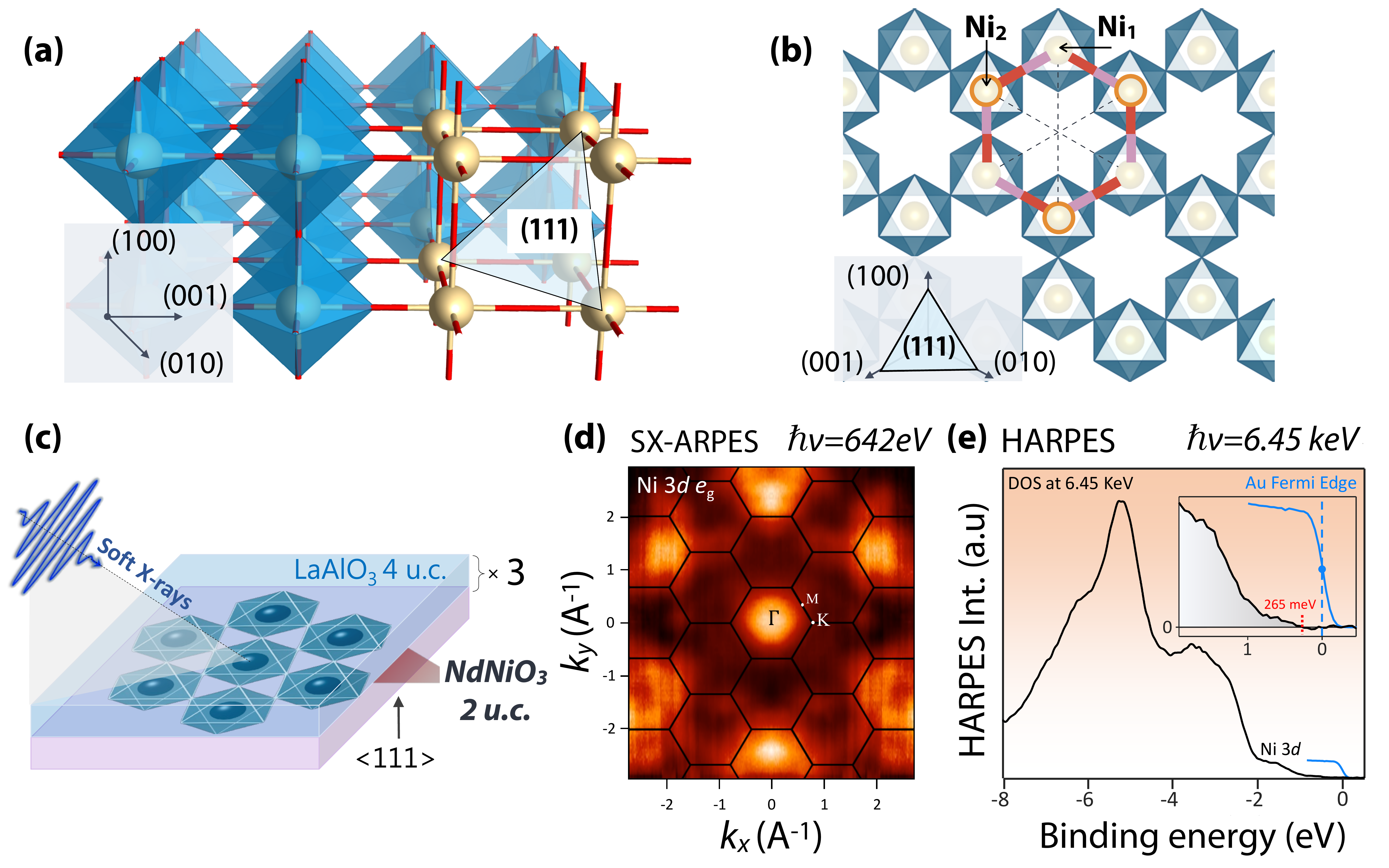}
\caption{\label{fig:LNO} (a) Two unit cells of an  infinite layer perovskite lattice  ABO$_3$ viewed along the $\langle 100 \rangle$ direction. A-site and oxygen atoms are omitted for clarity. (b) The same structure when grown along   $\langle 111 \rangle$ forms a graphene-like lattice with two sub-lattices and a buckled O-B-O bond. (c) Schematic diagram of the soft x-ray angle resolved photoemission spectroscopy (SX--ARPES) and hard x-ray photoemission spectroscopy (HAXPES) experimental geometry and the sample structure.  (d) Momentum resolved SX--ARPES photoemission intensity map for LaNiO$_3$ of the Ni 3\textit{d} states near the valence-band maximum measured with the photon energy of 642 eV.  (e) Bulk-sensitive HAXPES spectrum recorded at a photon energy of 6.45 keV with the estimated probing depth of approximately 85 \AA. Inset shows a high-statistics spectrum of the valence-band maximum (at the binding energy of 265 meV), referenced to the Au Fermi edge (Figures (c)-(e) adapted from Ref.[\onlinecite{Arab_nanolett_2019}]).}
\end{center}
\end{figure}

Conventionally, perovskite films and heterostructures are synthesized along the pseudo-cubic [001] direction. As seen in Fig.\ref{fig:LNO}, films grown along the [001] direction consist of alternating AO and BO$_2$ atomic planes. The same perovskite viewed along the [111] direction exhibits alternating AO$_3$ and B planes. Interestingly, by growing \textit{two} pseudo-cubic unit cells of ABO$_3$ along [111], one can generate an entirely new type of lattice with two vertically shifted triangular planes of B sites (see Fig. 1(b)). This artificially assembled buckled honeycomb lattice provides a unique opportunity to explore new phenomena due to the  superposition of complex $d$-orbitals in a graphene-like setting. 

What  is the reason for  such a dramatic change in the many-body  ground state caused by the new lattice motif? To start, we note that both tight-binding models and first-principles calculations indicate that $d$-electrons constitute the majority of states at the Fermi level\cite{Xiao:nc11,Ruegg11_2,Ruegg:prb12,Ruegg_Top:prb13,Yang:prb11a,Doennig:prb14,Wang:prb15}.  Here we focus on the partially filled shells with $e_g$ electrons. Unlike [001] oriented  perovskites that  strongly  favor real  orbitals, in [111] $d$-orbitals  posses $d$-wave symmetry (e.g. $|d\pm id\rangle = (|d_{z^2}\rangle \pm i |d_{x^2-y^2}\rangle)/\sqrt2)$ in the orbital  space, which in a sufficiently  weakly interacting  regime  also favor complex orbital orderings \cite{Ruegg11_2}.  In addition, for quarter-filing (e.g. low-spin 3\textit{d}$^7$), the Fermi level lies right at the $\Gamma$-point where two bands touch quadratically forming a so called ``quadratic band crossing'' (QBC) point \cite{Xiao:nc11,Ruegg11_2,Ruegg:prb12}. Moreover, \textit{\textit{  six-fold symmetry}} at the  QBC point protects it against   splitting into Dirac points\cite{Sun:prl09}. Finally, since in the $e_g$ manifold orbital  angular momentum is quenched, the  spin-orbit interaction only enters via mixing with $t_{2g}$-orbitals in the higher-order terms \cite{Stamokostas:prb18}. These features (the orbital structure of the states, along with the electronic band structure) set the stage for non-trivial theoretical possibilities.

Specifically, due to the buckled graphene-like  periodic arrangement of atoms,  theoretical calculations  for [111]-oriented bilayers of nickelates have predicted the realization of several exotic phases unattainable in either bulk or [001]-oriented heterojunctions \cite{Xiao:nc11,Ruegg11_2,Ruegg:prb12,Ruegg_Top:prb13,Yang:prb11a,Doennig:prb14,Wang:prb15,Okamoto_JPSJ}. For instance, model Hamiltonian calculations in the strongly correlated limit predicted that orbital ordering wins spontaneously over the bulk-like charge-ordered phase, whereas in the weakly correlated limit, a number of topological phases, including the Dirac half-metal phase, the quantum anomalous Hall insulator phase, and spin nematic phase with weak ferromagnetism, all  \textit{driven  by interactions without explicitly large spin-orbit coupling}, are feasible.  \cite{Xiao:nc11,Ruegg11_2,Ruegg:prb12}.   The details of these theoretical possibilities have appeared earlier in Ref.[\onlinecite{Fiete:jap15}].

Despite numerous predictions of interesting electronic and topological phases, the number of experimentally realized [111] lattice structures is  very limited \cite{Liu:mrs16}. To understand this situation, first we remind the reader that the majority of popular perovskite substrates (e.g. SrTiO$_3$, LaAlO$_3$, NdGaO$_3$, and YAlO$_3$) are polar along [111] with alternating +4e/-4e (or +3e/-3e) charges per unit cell for each atomic plane.  The large surface charges present a challenge, particularly because the process of initial nucleation and epitaxy on such high energy surfaces is  not well understood. In addition,  [111]  grown bi-layers of perovskites readily demonstrate  complex combinations of structural, chemical, and electronic reconstructions, which necessarily develop to compensate for the large electric fields  from the polarity jump \cite{Liu:mrs16, Rijnders_APL,Guo_APL_2018}. It is important to  emphasize that  unlike bulk crystals or thicker films, these imperfections in morphology and electronic structure are naturally amplified in a graphene-like  monolayer of the material.  The presence of top and bottom interfaces which are epitaxially coupled to the active monolayer constitutes another often poorly-understood control parameter.  

To shed light on those issues, monolayer-by-monolayer growth has been investigated by monitoring  \textit{in-situ} reflection high-energy electron diffraction (RHEED) during the growth progression of LaNiO$_3$ on SrTiO$_3$ (a polar interface) and  LaNiO$_3$  on LaAlO$_3$ (a non-polar interface) for the case of [111]-oriented substrates \cite{Middey:sr14}. The results revealed that in the polar case a non-perovskite La$_2$Ni$_2$O$_5$  phase rapidly develops within the first five unit cells followed by a gradual recovery  of  the desired LaNiO$_3$ phase. In sharp contrast, high-quality stoichiometric [111] LaNiO$_3$ was successfully stabilized on the LaAlO$_3$ [111] surface; these findings imply that a key to successful  [111] growth  is in a judicious choice of a film/substrate combination to avoid the polar mismatch at the interface \cite{Middey:apl12}.

In close connection, we can conjecture that  a few unit cells of a metallic or semiconducting buffer grown next to  the ionic polar surface (e.g. SrRuO$_3$, SrVO$_3$) may rapidly screen those charges and internal electric  fields to enable layer-by-layer synthesis even in the highly  polar cases. This attractive approach, however, may cause  ambiguity in the detection and interpretation of emerging metallic states in thin film against a trivial contribution from the conducting buffer layer.   

The experimental determination of the electronic, magnetic and structural properties of monolayer thick materials is another demanding venue.   Following theoretical   predictions~\cite{Xiao:nc11,Ruegg11_2,Ruegg:prb12,Ruegg_Top:prb13,Yang:prb11a,Doennig:prb14,Wang:prb15}, the behavior of 3$d^7$ electrons  on the buckled honeycomb lattice was investigated by resonant soft x-ray  absorption on a high quality [111] bilayer of   NdNiO$_3 $ capped by 4 unit cells LaAlO$_3$ templated on a LaAlO$_3$ [111] substrate. A detailed analysis of the angular dependent linear  dichroism  revealed  the presence of  a new ground state characterized by  antiferromagnetic correlations and  orbital ordering, unattainable in either bulk NdNiO$_3$ or in analogous heterostructures layered along the conventional (001) direction~\cite{Middey:2014p1570}.   

Furthermore, electronic band structure characterization by  Angle Resolved PhotoEmission Spectroscopy (ARPES) of an  active buried monolayer  presents an immense technical challenge and yet is highly desired because it is the most direct probe of topological band structures. The main obstacle standing in the way of such measurements is the fact that conventional probes of electronic structure including  ARPES or scanning tunneling spectro-microscopy (STM) are severely limited in their applicability to such systems due to  the high surface-sensitivity. 

Motivated by these challenges, novel bulk- and buried-layer-sensitive spectroscopic and scattering techniques, such as soft x-ray angle resolved photoemission spectroscopy (SX-ARPES) \cite{Gray_2013,Strocov_SRN} and hard x-ray photoemission spectroscopy (HAXPES) \cite{FADLEY200524,Gray_NM}, have recently emerged as viable probes for comprehensive investigations of ultrathin buried layers and interfaces due to their enhanced probing depth \cite{Jablonski_ARPES}.  

For instance, recently, Arab \textit{et al.} applied a combination of SX-ARPES and HAXPES to obtain the momentum-resolved and angle-integrated valence-band electronic structure of an ultrathin buckled graphene-like layer of [111] NdNiO$_3$ confined between two 4-unit cell-thick layers of insulating LaAlO$_3$ (see Fig.\ref{fig:LNO}c) \cite{Arab_nanolett_2019}. The direct measurements of the momentum-resolved electronic dispersion of the near-Fermi-level Ni 3\textit{d} valence states via SX-ARPES (shown in Fig.\ref{fig:LNO}d) provided an unambiguous evidence of such antiferro-orbital order and revealed a $P1$ structural symmetry arrangement in a 1$\times$1 unit cell, consistent with the results suggested by the x-ray linear dichroism (XLD) data \cite{Middey:2014p1570}. Complementary angle-integrated HAXPES measurements of the bulk-sensitive valence-band density of states (shown in Fig.\ref{fig:LNO}e) revealed the presence of a small bandgap of 265 meV, consistent with theoretical predictions. These early results suggest an effective strategy for investigating engineered states of matter even in buried structures of only a few atomic layers thick.

The rather thorough experimental investigation of [111] grown bilayers of  NdNiO$_3$ did not provide  evidence for a non-trivial topological state, the most likely of which would be a QAHE state \cite{Ruegg:prb12,Ruegg_Top:prb13}.  The null experimental result for the topological phase detection is somewhat surprising given that even the inclusion of realistic lattice distortions, such as the rotations of octahedral cages of oxygens surrounding the Ni ions, were included in the calculations and showed  no important effect on the presence  of topological states \cite{Ruegg_Top:prb13}.

Our best assessment of why the QAHE state was not realized is that the phase requires a rather fine balance of the on-site interaction energy and Hund's exchange coupling  on the Ni atoms  \cite{Ruegg:prb12,Ruegg_Top:prb13}.  The precise values of these quantities are difficult to determine since they depend in detail on the way the system is modeled.  In particular, theoretical models typically use a ``down folding" scheme in which higher-energy bands are ``integrated out" but taken into account in the low-energy models with effective on-site interaction and Hund's coupling \cite{Kotliar:rmp06}.  The higher bands result in a screening effect that reduces the effective interaction and exchange values (which are the ones appearing in a few band model) compared to their ``bare" values in a model that includes all the bands from all the atoms.  The predicted topological phase lies in a rather narrow range of the on-site interaction and Hund's coupling, and the material may happen to fall out of this range.  Another possibility is that the relatively small band gap predicted by theory is overwhelmed by disorder effects in the bilayer systems and magnetic domains form instead of a pristine topological state.

\begin{figure}[t]
\includegraphics[width=0.4\textwidth]{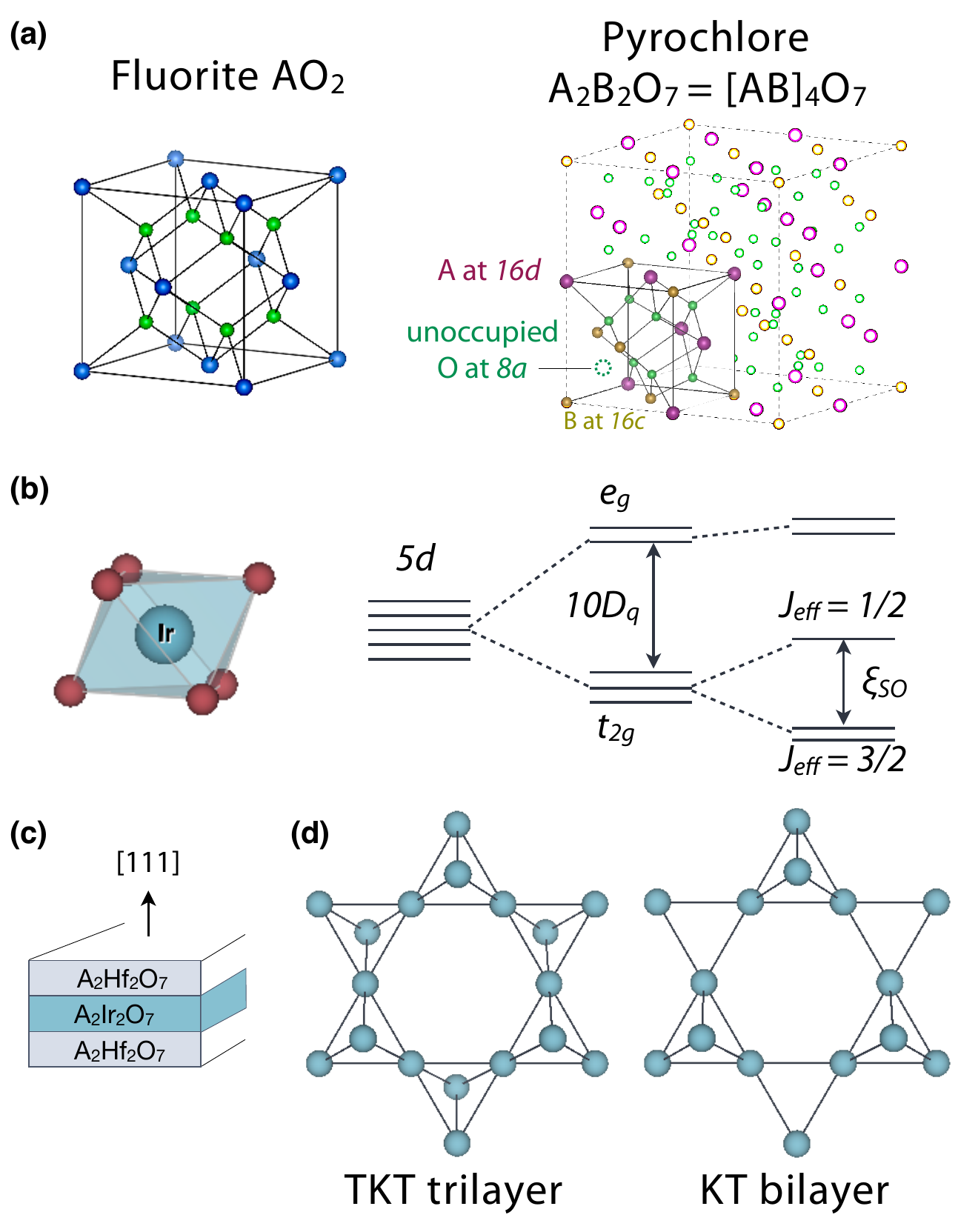}
\caption{\label{fig:pyro} (a) Lattice of the pyrochlore materials considered in this Perspective, exhibiting close resemblance to the fluorite structure. (b) Crystal field environment for A$_2$Ir$_2$O$_7$. (b) The Ir atom sits inside an octahedral cage of oxygen ions  O$^{2-}$ which results  in a splitting  10$D_q$ of the $d$-levels into an upper  $e_g$ manifold and a lower  $t_{2g}$ manifold. Spin-orbit coupling  $\xi_{SO}$ further splits  the $t_{2g}$ levels into $J_{\rm eff}=3/2$ and $J_{\rm eff}=1/2$ states.  For many pyrochlore compounds, the $J_{\rm eff}=1/2$ states are half-filled. If the spin-orbit coupling is large enough, one may  retain only these states in a low-energy model for a qualitative description.  (c) A sandwich structure of a [111] oriented film showing a thin layer capped by two insulators A$_2$Hf$_2$O$_7$ with a large band gap. (d)\ The alternating triangular-kagome-triangular (TKT), trilayer (T)\ and kagome-triangular (KT) bilayer lattice structure viewed along the [111] direction. These geometrically frustrated lattices can host multiple  emergent correlated and topological electronic states. }
\end{figure}

\section{[111] pyrochlore lattice}
\label{sec:pyrochlore}
While the ABO$_3$ perovskite systems are interesting in their own right, the A$_2$B$_2$O$_7$ pyrochlore systems bring in an addition element in the form of geometrical frustration. In these materials, the B sites form a pyrochlore lattice - a network of corner sharing tetrahedra.  As   seen in Fig.\ref{fig:pyro}(d), when viewed along the [111] direction the lattice consists of alternating kagome and triangular lattice planes.  Both the kagome and triangular lattices give rise to the phenomenon of  magnetic frustration, which can result in exotic fractionalized states with  quantum numbers of the electron (such as spin and charge)  broken apart or broken down into smaller values \cite{Maciejko:np15,Stern:arcmp16,Balents:nat10}.

The bulk crystal pyrochlore systems have gathered significant attention from  both the magnetism  \cite{Hozoi:prb14,Gardner:rmp10,Wang:prb17,Harter:sci17} and the correlated topological insulator communities \cite{Yang:prb11b,Yang:prb10b,Zhang:prl17,Guo:prl09,Kargarian:prb11,Schaffer:rpp16,Li:nc16,Witczak:prb12,Wan:prb11,Maciejko:prl14,Maciejko:np15,Pesin:np10}.  Of particular interest to this Perspective are the pyrochlore iridates,  A$_2$Ir$_2$O$_7$.  In this compound the Ir atom has an outer $d$-shell with 5 electrons, and  sits in an approximately cubic environment, leading to a splitting of the 5$d$ orbital into a subset of lower-lying $t_{2g}$ orbitals and higher lying $e_g$ orbitals, as depicted in Fig.\ref{fig:pyro}(b).  For heavy elements, like Ir, the spin-orbit coupling is sufficiently  strong  to lock the spin and angular momentum together producing an effective total angular moment $J_{\rm eff}=3/2$ and $J_{\rm eff}=1/2$ , though there may be some degree of mixing when the spin-orbit coupling is large\cite{Stamokostas:prb18}.  Five electrons in the $d$-shell thus corresponds to a half-filled $J_{\rm eff}=1/2$ state, and parallels can be made to spin 1/2 systems. This feature entwined  with  the frustrated lattice is a major motivator for theoretical studies of the pyrochlore  iridates and osmates.

An important general consideration for topological states is the spatial dimension of the system.  Certain topological states, such as Weyl semimetals exist in three spatial dimensions, but not two \cite{Armitage:rmp18}. Thus, the film thickness is an important parameter in the realization of topological states.  In the ultra-thin limit, the system is effectively two-dimensional and states such as the QAHE (a zero magnetic field quantum Hall state) are possible.  For spin systems (local moments on the lattice), the spatial dimension is also important--the lower the dimension, the larger the quantum fluctuations and hence the easier  to achieve even more exotic magnetic ground states such as quantum spin liquids \cite{Balents:nat10,Wen}.  

It is important to note that the vast amount of literature over the last 15 years on topological band structures is largely a subset of what are now referred to more generally as ``symmetry protected topological" states or SPT states.  For example, time-reversal invariant topological insulators with a non-trivial $Z_2$ invariant require the imposition of time-reversal symmetry to provide a sharp distinction from trivial band insulators with a trivial $Z_2$ invariant \cite{Hasan:rmp10,Moore:nat10,Qi:rmp11}.  Topological crystalline insulators require some additional lattice symmetries, such as a mirror symmetry in the lattice that gives rise to a mirror Chern number, to provide a sharp distinction between the topological and non-topological phases \cite{Fu:prl11,Hsieh:naC2012}.  There have also been studies of non-symmorphic space groups \cite{Liu14,Shiozaki:prb16}, and many other discrete symmetries \cite{Schnyder:prb08,Ryu:njp10}.

The three-dimensional Weyl and Dirac systems, which have garnered much interest lately also require certain symmetries to be present for their realization and/or stability.  Specifically, the Dirac metals require a four-dimensional irreducible representation of the small groups at specific momenta (for symmetry protection)\cite{Young:prl12,Steinberg:prl14}, and the Weyl systems entail either inversion symmetry or time-reversal symmetry breaking \cite{Armitage:rmp18}.  Another important way a Dirac metal can occur is at the phase boundary between a time-reversal invariant topological insulator and a trivial insulator \cite{Fu:prl07,Murakami:prb07}.  These topological semimetals and their other variants, such as nodal line semimetals, have been reviewed recently \cite{Armitage:rmp18,Burkov:nm16}. 

On the experimental front, the ability to grow such pyrochlore systems in thin-film form along the [111] direction offers opportunities to ``design" interacting topological materials.  In particular, {\em en route} to the two-dimensional limit from the three-dimensional limit, one may encounter ``hidden" topological states \cite{Yang:prl14} in the electronic band structure.  There is now a rather vast literature on predicted topological states in [111] grown pyrochlore iridate thin films \cite{Hu:prb12,Hu:sr15,Hwang:sr16,Yang:prl14,Chen_DMFT:prb15,Laurell:prl17}, and other related systems \cite{Koksal:prb18,Ruegg:prb12,Ruegg_Top:prb13,Yang:prb11a,Okamoto:prb14,Liang:njp13,Lado:prb13,Doennig:prb14,Ruegg11_2,Wang:prb15,Xiao:nc11,Fiete:jap15,Oja:prb08,Okamoto:prl13,Bing:prb14,Moreau:prb17,Cossu:epl17,Tahini:prb16}.  As we mentioned in the introduction, it is not our intent to provide a comprehensive review, but rather highlight the richness of possibilities in this system.

The fabrication of A$_2$Ir$_2$O$_7$ thin films and heterostructures remains extremely challenging due to the existence of several closely related critical issues. To understand them, it is necessary to emphasize the structural and chemical peculiarities of the pyrochlore iridates. First, pyrochlore oxides host geometrically frustrated interpenetrating A and B sublattices, such that the pyrochlore structure can be regarded as an anion-deficient fluorite structure (A$_2$B$_2$O$_7$ = [AB]$_4$O$_7$), whose A and B cations alternatively occupy the fluorite cation sites along the <110> direction, as illustrated in Fig. 2a. The anion vacancies reside in the tetrahedral interstices between adjacent B-site cations. As a result, defects can naturally enter into the structure, forming off-stoichiometric phases A$_{2\pm x}$B$_{2\pm y}$O$_{7\pm \delta}$. Unfortunately, the electronic properties of pyrochlore iridates are very sensitive to the non-stoichiometry; For example, in Eu$_2$Ir$_2$O$_7$, approximately $\sim$4\% of extra Ir  is sufficient to entirely suppress the metal-insulator transition \cite{Ishikawa_PRB_2012}.

Another formidable issue lies in the oxidation of iridium, which forms  highly volatile gaseous iridium oxides (e.g. IrO$_3$) during the growth. This elicits  a dramatic loss of Ir inside the film \cite{kim_JAP_2008,Jehn_PMR_1978}. The inherent Ir volatility acts to strongly magnify the formation of defects, further deviating the structure  from the pyrochlore phase towards  multiple secondary chemical phases. Here we emphasize that these issues are inherent to growth with all platinum metals, including gaseous phases of platinum, rhodium, iridium, ruthenium and osmium oxides \cite{Chaston_PMR_1965}.  

Therefore, to stabilize high quality films, one may naturally attempt to grow the films in high vacuum and at low temperature. However, this in turn triggers two other problems. First, due to the low reactivity of Ir metal, Ir$^{4+}$ oxidation can only be achieved with great difficulty in the vacuum environment. Secondly, growing materials at low temperatures yields samples with very poor morphological quality.      

With the growth  challenges described above, layer-by-layer growth of [111]-oriented pyrochlore iridate ultra-thin films  has yet to be demonstrated. Nevertheless, significant advances in the fabrication of [111] A$_2$Ir$_2$O$_7$ thin films (A = Nd \cite{Gallagher_SciRep_2016,Kim_PRB_2018}, Eu \cite{Fujita_SciRep_2015,Fujita_APL_2016}, Tb \cite{Kozuka_PRB_2017,Fujita_PRMater_2018}, Pr \cite{Ohtsuki_PNAS_2019,Guo_arXiv_2019,Ohtsuki_JAP_2020}, Bi \cite{Yang_SciRep_2017,Yang_PRMater_2018,Chu_NJP_2019}) and heterostructures \cite{Fujita_PRB_2016} have been steadily demonstrated. 

The vast majority of the successfully grown A$_2$Ir$_2$O$_7$ thin films have been realized via the so-called ``solid phase epitaxy''. In this method an amorphous film of the proper stoichiometry is first deposited on a [111]-oriented yttria-stabilized zirconia (YSZ) substrate by pulsed laser deposition or sputtering at reduced temperature and under low-oxygen partial pressure of pure O$_2$ or Ar/O$_2$ mixture, followed by\textit{ ex-situ} post-annealing at elevated temperatures to crystallize the amorphous film into the desired pyrochlore structure. Such a two-step growth protocol allows for the fabrication of reasonably high-quality A$_2$Ir$_2$O$_7$ films down to a few tens of nanometers albeit with mediocre surface roughness. It should be noted that the growth window for each specific compound is rather narrow, lacks universality, and requires dedicated optimization \cite{Fujita_SciRep_2015,Gallagher_SciRep_2016,Yang_SciRep_2017}.

It is noteworthy that the films obtained by the ``solid phase epitaxy'' technique exhibit distinctly different transport behaviors from their bulk counterparts. In the bulk most of the  A$_2$Ir$_2$O$_7$ compounds show a metal-insulator transition (MIT) (except for A = Pr and Bi, which are pure metals) \cite{WK_ARCMP_2014}. Concurrent with the MIT, the paramagnetic phase transits into an antiferromagnetic phase with the peculiar all-in-all-out (AIAO) spin configuration stabilized on each cation sublattice. Such a spin configuration has two degenerate domain structures -- all-in-all-out (AIAO) and all-out-all-in (AOAI) -- that are `switchable' by time-reversal operation \cite{Arima_JPSJ_2013}.        

Strikingly,  not found in bulk, an unusual odd-parity field-dependent term in the magnetoresistance, together with a zero-field offset in the Hall resistance were lately discovered in [111] Eu$_2$Ir$_2$O$_7$ thin films below the onset of the MIT \cite{Fujita_SciRep_2015}. Since the Eu$^{3+}$ ion is non-magnetic, the observed exotic phenomena  directly reflect the magneto-transport response of the carriers coupled to the localized moments of the Ir sublattice. Specifically, it was argued to originate from the formation of the single  AIAO domain defined by the direction of the cooling magnetic field. The observed magnetic domains are exceptionally rigid and once established, switching of domains can no longer be realized by sweeping experimentally accessible external fields \cite{Fujita_APL_2016}.

On the other hand,  in pyrochore iridates with \textit{magnetic} A$^{3+}$ ions, because of the strong magnetic coupling between 4$f$ and 5$d$ moments, switching of the Ir domains is easily accomplished by sweeping the external magnetic field along the [111] direction \cite{Ueda_PRL}. The proposed mechanism of domain switching has been captured in [111] Tb$_2$Ir$_2$O$_7$ thin films \cite{Kozuka_PRB_2017,Fujita_PRMater_2018}. Interestingly, the effect of domain switching  leads to the formation of domain walls in the \{111\} planes, where two-fold rotational symmetry can be broken. As the result, an intrinsic anomalous Hall conductance due to non-zero Berry curvature may be probed at the domain walls. This phenomenon was recently demonstrated in [111] Nd$_2$Ir$_2$O$_7$ thin films \cite{Kim_PRB_2018}, as well as at the [111] heterointerface between two pyrochlore iridates, Eu$_2$Ir$_2$O$_7$ and Tb$_2$Ir$_2$O$_7$ \cite{Fujita_PRB_2016}.                 

In addition, for all metallic A$_2$Ir$_2$O$_7$, a spontaneous Hall effect (i.e. anomalous Hall effect without net magnetization) due to the spin chirality associated with Pr 4$f$ moments was previously reported in bulk Pr$_2$Ir$_2$O$_7$ below 1.5 K  \cite{Machida_Nature_2010}.  Surprisingly, the onset of the spontaneous Hall effect has been recently reported to appear at 50 K \cite{Ohtsuki_PNAS_2019} or 15 K \cite{Guo_arXiv_2019} in [111] Pr$_2$Ir$_2$O$_7$ thin films -- a temperature much higher than the spin-correlation scale due to Pr moments. Based on these results, the enhancement was speculated  to originate from the Ir 5$d$ moments. One argument given by Ohtsuki \textit{et al.} suggests that [111] epitaxial strain may induce the AIAO magnetic order on the Ir sublattice causing  a magnetic Weyl semimetal state to appear in the film \cite{Ohtsuki_PNAS_2019}. However, as no magnetic reflections associated with the AIAO order were detected by resonant scattering, Guo \textit{et al.} argued that the spontaneous magnetization is  likely due to the localization of the Ir moments, which either creates additional spin chirality or strengthens the effective Pr-Pr coupling via the 4$f$ - 5$d$ interaction \cite{Guo_arXiv_2019}. Additionally, linear magneto-resistance (MR) linked to the multiple types of charge carriers was found in [111] Bi$_2$Ir$_2$O$_7$ thin films, exhibiting striking resemblances to the scale invariant MR in the strange metal state of high Tc cuprates \cite{Yang_SciRep_2017,Yang_PRMater_2018,Chu_NJP_2019}.    

At  this  point, the conclusion we  draw  from these findings in A$_2$Ir$_2$O$_7$ films is that the magnetic properties are significantly more complex than those of  the corresponding bulk systems.  The relative abundance of transport data can be   contrasted to  the challenges of performing other kinds of advanced measurements.  For instance, magnetic neutron scattering is difficult  because of the small number of scatterers  to get a strong scattering signal and the fact that Ir is a strong neutron absorber \cite{Choi:prl12}.   The application of ARPES and STM  is challenging due to the problem of \textit{in-situ} preparing clean and atomically smooth surfaces of pyrochlore films. Furthermore, the potentially rich physics of \textit{collective excitations} in thin films of Ir pyrochlores is almost entirely unexplored. This is primarily  due to    the low  sample volume (e.g. for inelastic neutrons) and lack of momentum information for probes with high energy resolution (e.g. optical  probes). One method that may prove to  be sufficiently powerful for these systems is the resonant inelastic x-ray scattering (RIXS) which shows a strong response for Ir \cite{Ament:rmp11}.

\section{Outlook} 
\label{sec:outlook}

In this short Perspective, we have highlighted some of the key theoretical ideas motivating the drive to produce high quality [111] transition metal oxide thin films. One major point is that the material conditions are ripe for exotic many-body quantum states with topological properties or fractionalized excitations.  On the experimental side, we have attempted to lay out some of the key materials fabrication challenges facing thin film growth.  We hope our discussion will help other researchers to grasp gaps in our current understanding and see them as  attractive opportunities to  explore.  

\begin{figure}[t]
\includegraphics[width=0.5\textwidth]{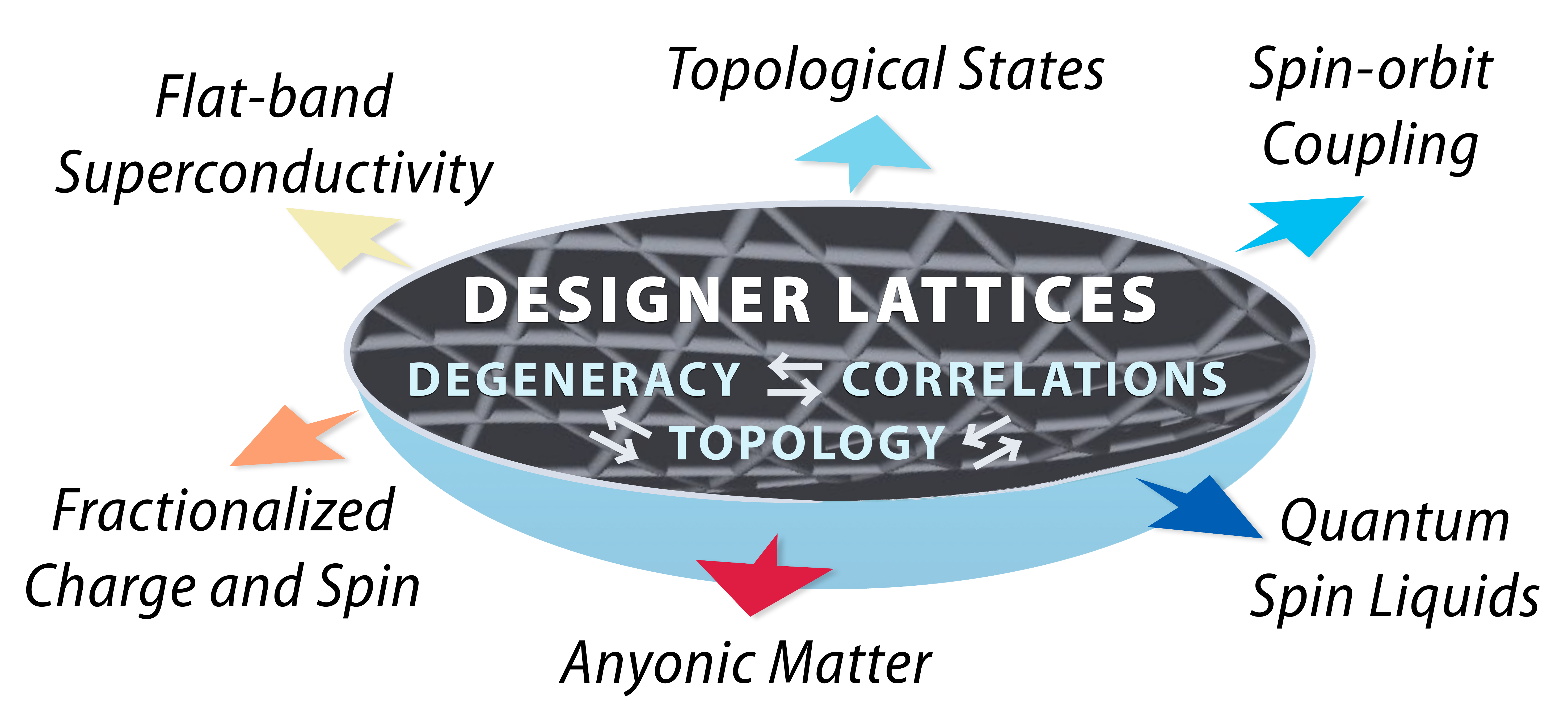}
\caption{\label{fig:future} Possible new quantum states and many-body phenomena that emerge on designer lattices, where the element of geometrical frustration is introduced.}
\end{figure}

We will conclude the Perspective by looking into a possible future. From the materials side, while the current focus is mainly on perovskite and pyrochlore [111] oxides with active $d$ electrons, systems dominated by $f$ electrons can potentially lead to striking electronic and magnetic properties, especially where the $d$-$f$ interaction is strong. In addition, this methodology can also be applied to structures with more interactive degrees of freedom \cite{Coleman_ARCMP}.

One of the promising candidates is the family  of  spinels AB$_2$X$_4$, in which both the A and B sites can be either magnetic or non-magnetic, and the X sites may be an element  from the chalcogen family (group 16 of the periodic table). In this structure [111] quasi-two-dimensional slab is  composed of alternating stacking of kagome and triangular cation planes with magnetic frustration, leading to the potentially  rich and exotic magnetism \cite{XL_Nanolett_2019}.  Following the same line of thought, we can  envision that  the  current scarcity of  high  quality Kitaev magnets can be mitigated  by the designer approach akin to the  graphene lattice derived from [111] perovskites. Finally, a large class of frustrated  compounds with tri-fold symmetry including jarosites, wurtzites, and garnets await to be  stabilized in the thin
film form and investigated  for  interesting quantum and  topological phases \cite{IFM_book}. 

All in all, as sketched in Fig. 3, designer lattices such as the kagome and other thin film systems with geometrical frustration discussed in this Perspective, offer a new  framework to explore some of the most interesting frontiers in contemporary condensed matter physics. Among those, the captivating possibilities of interacting topological states, quantum spin liquids, anyonic matter, fractionalized charge and spin fermions, and flat band superconductivity still  lie in wait for their realization.

\begin{acknowledgments}
The authors deeply acknowledge D. Khomskii, P. Coleman, P. Chandra, M. Kareev, J. Freeland and J. Pixley for insightful discussions. JC and XL were supported by the Gordon and Betty Moore Foundation's EPiQS Initiative through grant GBMF4534 and by the Department of Energy under grant DE-SC0012375. GAF was supported by NSF Grant No. DMR-1949701,  NSF Materials Research Science and Engineering Center Grant No. DMR-1720595, and a QuantEmX grant GBMF5305 from ICAM and the Gordon and Betty Moore Foundation.
\end{acknowledgments}

\bibliography{pyro111_Jak}

\end{document}